\documentclass[preprint2]{aastex}

\shorttitle{THE INFLUENCE OF BULGE PROFILE SHAPES}
\shortauthors{GRAHAM \& PRIETO}

\begin{document}

\input{psfig}

\title{THE INFLUENCE OF BULGE PROFILE SHAPES ON CLAIMS FOR A SCALE-FREE HUBBLE SEQUENCE FOR SPIRAL GALAXIES}
\author{Alister W. Graham\altaffilmark{1} and Mercedes Prieto}
\affil{Instituto de Astrof\'{i}sica de Canarias, La Laguna, 
E-38200, Tenerife, Spain}

\altaffiltext{1}{Isaac Newton Group, La Palma, Spain\\agraham@ll.iac.es, mpm@ll.iac.es}

\begin{abstract}
We investigate recent claims that the Hubble sequence of spiral galaxies 
is scale-free.  Fundamental to this investigation is the fact that within 
the photometric data of 86 spiral galaxies from de Jong \& van der Kruit 
(1994) - from which these claims were made - a trend exists between 
morphological type and bulge profile shape.  While late-type spiral bulges 
are described by an exponential luminosity profile, the early-type
spiral bulges are better described by an $r^{1/2}$ or $r^{1/4}$ law.
Taking the scale-lengths from the best-fitting surface brightness 
profile models (i.e.\ either using exponential, $r^{1/2}$, or $r^{1/4}$ 
law profile parameters), we show that in all six passbands used (BVRIHK) the
early-type spirals have a larger $r_e/h$ ratio than late-type spiral galaxies.
In contrast to this, fitting an exponential profile to the
bulges of all spirals results in the mean $r_e/h$ ratio for the early-type
spiral galaxies actually being smaller than the mean $r_e/h$ ratio for the late-type
spiral galaxies (at the 3$\sigma$ significance level using K-band data).
%
%
\end{abstract}


\notetoeditor{{\bf ONE:} I was not able to get the  slash-email{ }  command 
to print our email addresses, so I have used the option you suggested.
{\bf TWO:} Table two still has problems - it overruns the pagelength.}

\section{Introduction}

De Jong (1996b) and Courteau, de Jong, \& Broeils (1996) have suggested that 
``the Hubble sequence of spirals is scale-free''.  They claim that  
``the constant ratio of bulge-to-disk scale-lengths appears to be independent 
of galaxy type''.   If true, this would not only be at odds with the 
classification scheme posed by Hubble (1926,1936) and later Sandage (1961) - 
in which the bulge-to-disk ratio progressively decreases as one goes from 
early to late type spirals (Simien \& de Vaucouleurs 1986) - but would have 
consequences for theories of galaxy formation. 

While the surface brightness profiles of the disks of spirals are well described
by exponential models, the light profiles of the bulges are known to possess 
a range of structural shapes (Andredakis, Peletier \& Balcells 1995;
Carollo, Stiavelli \& Mack 1998).  These can be easily modelled with
the Sersic (1968) $r^{1/n}$ law.  A generalization of de Vaucouleurs'
(1948) $r^{1/4}$ law, the free parameter $n$ can describe the observed range
of bulge profile shapes.  Indeed, a subset
of this model (namely $n$=1, 2, 4) was applied by de Jong (1996b) to
the surface brightness profiles of the bulges in his sample of 86 face-on 
spiral galaxies. 
de Jong found that 60\% of his sample were better modelled (based on the
$\chi ^2$ statistic) with an $n$=1 model, while 40\% preferred a
larger value of $n$, with 15\% preferring $n$=4.  Similarly,
Courteau (1996) found, when fitting both an $n$=1 and an
$n$=4 profile model, that 15\% of the bulges in his sample of spiral 
galaxies were better modelled with the $n$=4 profile 
(Courteau et al.\ 1996). 

Given these results, Courteau et al.\ (1996) reported 
that late-type spirals are best fitted by two exponential models,
and they chose to represent all their spiral galaxies this way.
Subsequently, their claim for a scale-free Hubble sequence for spirals 
was based upon structural parameters obtained from fitting exponential 
light profile models to both the disk and the bulge.

However, the above percentages become most interesting when one notes 
that the galaxies preferring the larger values of $n$ are the early-type
spirals, while the late-type spirals prefer values of $n$$\sim$1 
(Andredakis et al.\ 1995; Moriondo, Giovanardi, \& Hunt 1998). 
Additionally, Andredakis et al.\ (1995) have shown 
that the bulge-to-disk ratio of luminosities
varies systematically with profile shape, such that galaxies with larger
bulge-to-disk luminosity ratio have larger shape parameters. 
Logically, any conclusions drawn from structural parameters which have 
ignored these structural differences must surely be questioned 
(Moriondo et al.\ 1998).  By using the best-fitting profile models
(either $n$$=$1, 2, or 4) this paper reinvestigates the claim for a 
scale-free Hubble sequence of spiral galaxies.

\section{Data}

We have re-analyzed the data presented by Courteau et al.\ (1996).
They presented two data sets, however, only one is appropriate for 
explorations of galaxy properties as a function of morphological type.

Lahav et al.\ (1995) showed that the dispersion in galaxy type index, T, 
between six 
experienced galaxy classifiers was on average 1.8 T-units, and 2.2 T-units
when comparing the RC3 (de Vaucouleurs et al.\ 1991) T-index with those of
the six classifiers.  A similar figure of disagreement (2.0-2.5 T-units) 
was obtained by four human classifiers of HST images (Odewahn et al.\ 1996). 
Unfortunately, because of this, the larger of the two data sets presented 
in Courteau et al.\ (1996) - 243 Sb--Sc galaxies from the 349 Sb--Sc 
galaxies of Courteau (1996) - can not in themselves be used to 
explore possible trends within the Hubble sequence of spiral galaxies.  

What the R-band data of Courteau (1996) does enable, is to show that 
the individual ratios of bulge-to-disk scale-lengths span a broad 
range of values (Courteau et al.\ 1996, Figure 1).  
Scale-length ratios within just one standard deviation of the median are
shown to span a range greater than a factor of 4, with the 1$\sigma $ 
confidence interval ranging from 0.029 to 0.135, and a long tail in the 
distribution stretching to 0.35.  
To obtain the ratio of the bulge effective radius $r_e$ to the disk
scale-height $h$, these numbers should be multiplied by 1.679, giving 
ratios of $r_e/h$ up to $\sim$0.6.  Therefore, in passing, we stress that
caution should be employed when using any sort of mean bulge-to-disk 
scale-length ratio, since a broad range of values 
spanning one order of magnitude exists amongst the real galaxy population.

The second data set, that of de Jong \& van der Kruit (1994), 
is however useful. 
It includes galaxy types from Sa through to Sm.  This sample of 86
galaxies actually includes two S0 galaxies which are removed here as 
de Jong (1996b) notes that their surface brightnesses are significantly
below the trend seen for the rest of the spiral galaxies, and their connection
with the early-type spiral galaxies is still unclear. 
The sole irregular galaxy (T=10) is 
also removed, leaving 83 face-on (minor over major axis ratios greater 
than 0.625) Sa to Sm galaxies, imaged in six passbands (BVRIHK).

\section{Analysis}

In recent years, some of the limitations of the classical surface brightness
profile models, such as the exponential or the $r^{1/4}$ law, have
been realised.  Departures in the radial falloff of light from these
models has been not only detected but successfully modelled for: 
the dwarf galaxy population (Davies et al.\ 1988; Young \& Currie 1994;
Binggeli \& Jerjen 1998),
the ellipticals (Caon et al.\ 1993; Graham \& Colless 1997), 
brightest cluster galaxies (Graham et al.\ 1996), and for the bulges
of spirals (Andredakis et al.\ 1995).  

The Sersic (1968) 
law has proved successful in parameterizing such departures from the 
traditional models and can be written as 
\begin{eqnarray}
I(r)&=&I_{0}\exp \left[ -\left( \frac{r}{h_b}\right) ^{1/n} \right] \nonumber \\
    &=&I_{e}\exp \left[ -(2n-0.327) \left\{ \left( \frac{r}{r_e} \right) ^{1/n} -1 \right\} \right] . \nonumber \\
\end{eqnarray}
The first line shows how the intensity $I$ varies with radius $r$; $I_{0}$ is
the central intensity where $r$$=$0.  We use $h_b$ here to denote the bulge scale-length
and avoid confusion with the disk scale-length which is denoted by $h$ elsewhere in
this paper.  The third model parameter, $n$, describes the level of curvature in the 
light profile.  For example, when $n$=1 the 
Sersic law is equivalent to an exponential light distribution; 
when $n$=4 it mimics the de Vaucouleurs $r^{1/4}$ law.  The value
of $n$ is of course not restricted to integer values and remains 
meaningful up until values of around 10-15.
The second line is a variant of the first expression, with the model parameters
now $I_{e}$, the intensity at the radius $r_{e}$ which encloses half of the 
total light of the bulge.  Equating like-terms, one has that 
$I_0$$=$$I_e\exp (2n-0.327)$ and $(r_e/h_b)$$=$$(2n-0.327)^n$.  Therefore, when
$n$$=$1, $r_e$$=$1.67$h_b$, and when $n$$=$2, $r_e$=13.5$h_b$.
One can also easily see why effective radii rather than scale-lengths are 
used for the $r^{1/4}$ law, since $h_b$=$r_e$/3466.  Given that this paper
uses parameters from $n$=1, 2, and 4 Sersic models, we have used 
effective radii rather than scale-lengths.

de Jong (1996a) fitted three models to the surface brightness profiles of the 
bulges, all with an accompanying exponential profile model to the disk.
The goodness-of-fit for each model was measured using the $\chi ^2$ 
statistic.\footnote{The data can be found at 
\url{http://cdsweb.u-strasbg.fr/htbin/Cat?J/A+AS/118/557}.}
For the B, V, H and K passbands, it is observed that for every
two galaxy bulges that are best fit with an $n$=2 or $n$=4 profile, 
there are three galaxy bulges whose best-fitting profile model is the 
$n$=1 model.  For the R and I passbands, the number of galaxy bulges 
best fit with the $n$=1 model equal the number of bulges better fitted 
with the alternative $n$=2 or $n$=4 models (Table~\ref{chi2-tab}). 

In using the best-fitting bulge models (either $n$$=$1, 2, or 4) 
the associated model parameters were not always reliable.  
In particular, the $r^{1/4}$ model sometimes resulted in 
values for $r_e$ that were either inaccurately determined 
and/or were unrealistically large.  To accommodate for this, each value 
of $r_e$ was inspected and the galaxy either retained, or rejected if 
$\Delta r_e/r_e$$>$40\% or $r_e/r_{max}$$>$0.5, 
where $r_{max}$ is the maximum radius for 
which the the surface brightness profiles extend ($\sim$26$\pm$1 in B). 
This typically resulted in the exclusion of only 1 or 2 galaxies from each of 
the morphological type bins T=1-3 and T=7-9 used in this comparative study.

Table~\ref{stats-tab} shows the difference in the mean value of 
$r_e/h$ for the early- and late-type morphological class bins used
by Courteau et al.\ (1996).  
It shows this ratio for the K- and R-band data fit 
with an exponential bulge model by de Jong (1996b) and 
Courteau et al.\ (1996).  Using the best-fitting $n$$=$1, 2, and 4
models, we present this difference 
of means for all six passbands (BVRIHK).  However, this difference 
in the ratio is meaningless on its own.  What is important is the 
significance of this 
difference, and this depends on the sample size and standard deviation
of the distributions.  To this end, we have applied Student's t-test.
The probability, Prob(t), that the difference in means could be as large
as it is by chance 
is given in Table~\ref{stats-tab}; small values indicate that the
means are significantly different from each other.

\section{Discussion}

The majority of the early-type spirals ($\leq$Sb) prefer to have 
values of $n$$>$1, while late-type galaxies ($\geq$Sd) are better 
fit with an exponential bulge (see Table~\ref{chi2-tab}).
%
%
The universal application of the exponential fitting function 
ignores from the start real differences in 
galaxy structure, and introduces a systematic bias into the parameterization
of these galaxies - under-estimating the effective 
half-light radius of the bulge.\footnote{A similar 
behaviour is known to occur with $r^{1/4}$ modelling of the light 
profiles of elliptical galaxies (Graham \& Colless 1997, Figure 11).}
Figure~\ref{4to1-fig} shows the ratio of the effective radii derived from 
the $r^{1/4}$ model ($r_{e,4}$) and the effective radii derived from the 
exponential model ($r_{e,exp}$), plotted against the ratio of the exponential 
model disk scale length co-fitted with the $r^{1/4}$ bulge model ($h_4$) 
and the exponential disk scale length co-fitted with the exponential bulge 
model ($h_{exp}$).  It shows that $r_{e,4}/r_{e,exp}$$>$1, while the exponential 
disk scale-length remains largely unchanged as the bulge profile model is adjusted. 

Similarly, fitting an $n$=1 profile will over-estimate 
the half-light radii for some of the late-type spirals. Although de Jong 
(1996a) shows for the late-type spirals that an $n$=1 model provides  
a better representation of the bulge than an
$n$=2 or $n$=4 model, he also notes that values as low as $n$=0.5 are
obtained when applying the Sersic profile to the bulge (de Jong 1996a). 
Furthermore, 
Andredakis et al.\ (1995), in fitting the Sersic model to the K-band
bulge light profiles of 30 spiral galaxies, found 
some Sb--Sd galaxies to have bulge profiles with shape
parameters smaller than 1.  Consequently, 
restricting the structural profiles of
all late-type galaxies to be described by an $n$=1 model may be 
increasing their mean bulge scale-length and hence 
reducing the true difference between the $r_e/h$ ratio of the 
early- and late-type spirals.  That is, the probabilities in Table~\ref{stats-tab}
may be larger than they should.


As stated by de Jong (1996b), K-band data is the passband of choice
for such studies, making it ``possible for the first time to trace 
fundamental parameters related to the luminous mass while hardly 
being hampered by the effects of dust and stellar populations.''
Indeed, some of the galaxies in de Jong sample were noted to possess 
dust lanes and circumnuclear star formation. 
Furthermore, bulges are brighter in K than in B with respect 
to the disk, and since the bulge/disk decomposition is easier when the 
bulge is relatively brighter, the fitting algorithm therefore works 
better in the K-band (de Jong 1996b).

Using exponential bulge models, Courteau et al.\ (1996) mention that 
the $r_e/h$ ratios of the early-type spirals appear systematically 
below the average $r_e/h$ value for all spiral galaxy types.  
They assert that this difference is not large, and claim that the 
constant ratio of bulge-to-disk scale-length is independent of Hubble type.  
However, our analysis of the exponential models fitted to the K band data 
of de Jong (1996b, Figure 18) reveals that the mean value of $r_e/h$ for the Sa--Sb
type galaxies is actually smaller than that for the late-type spirals at
a significance of 98\% (3$\sigma$)! (Table~\ref{stats-tab}).  
%
%
Similarly, with the R-band data presented by Courteau et al.\ (1996), 
and in fact for all wavelengths used (excluding the V-band),
the ratio of $r_e/h$ is smaller for the Sa--Sb galaxies than it is for 
types $\geq$Sbc.    This result is at odd with the
classical picture of the Hubble sequence, where early-type spirals 
have larger bulge-to-disk scale-length ratios than late-type spirals.

Due to the use of exponential bulge models for 
the Sa--Sb type galaxies, the above result can be understood in
terms of systematically under-estimating the size of these bulges. 
Correcting for this, by taking the best-fitting 
structural parameters, from either the $n$=1, 2, or 4 models,
we find that the situation reverses itself.  The average value of $r_e/h$
for the Sa--Sb type galaxies is found to be larger than the average
value of $r_e/h$ for galaxy types $\geq$Sd, in all six passbands.
Table~\ref{stats-tab} shows that the probability that the Sa--Sb 
type galaxies have the same mean $r_e/h$ as the Sd--Sm type galaxies is 
weakly ruled out at the 1.5--2 $\sigma$ level in five of the six passbands 
used by de Jong \& van der Kruit (1994).
Interestingly, it is the K-band data which suggest that the difference in 
means is not significant.  However, this result in itself is significant
when compared to the result obtained using only exponential bulge profile 
models.  Using the best-fitting models, the average $r_e/h$ ratio for the 
sample is larger -- at the 3 $\sigma$ significance level -- than when 
obtained using only the $n$=1 model.

We plan to refine this work by fitting a Sersic profile 
with free (i.e.\ not fixed) shape parameter, $n$, 
to the bulges of the spirals in the sample of de Jong \& van der
Kruit (1994).  Furthermore, 
Courteau et al.\ (1996) noted that about 1/3 of the sample of galaxies 
from de Jong (1996a) had a bar modelled as an additional component -
requiring eight structural model parameters for these galaxies.   While 
de Jong modelled a bar when fitting the exponential bulge models to the 
2D images, his one-dimensional decomposition technique which he used to fit the
$r^{1/4}$ and $r^{1/2}$ bulge models did not allow for the influence of a bar.
Subsequently, we must caution that failure to model the bar in the 1D 
data used here may influence the scale-lengths obtained.


%
Arguments for secular evolution, namely the exponential bulge light 
profile and the restricted range of bulge-to-disk scale-lengths, 
are either wrong or questionable. 
Andredakis et al.'s (1995) alternative to secular evolution - based upon 
the continuous trend between galaxy structure, as measured by $n$, and 
galaxy type - 
is largely supported by the data of Courteau et al.\ (1996).
In the framework of this model, n-body simulations (Andredakis 1998) have
shown how the imprint of disk formation is left upon the bulge, 
creating the observed trend between shape parameter and morphological type.
Yet another alternative is offered by Aguerri \& Balcells (1999), 
where the shape of the bulge grows from an $n$=1 profile to larger 
values of $n$ as shown through n-body simulations of merger events. 

Whether the bulges of spiral galaxies formed after the disk, as in the 
secular evolution model (Courteau et al.\ 1996), or, whether the bulge 
is in fact older 
than the disk (Andredakis 1998, and references within) may be better 
answered when the range and trends of bulge-to-disk ratios are 
better known.

\acknowledgments

We thank Marc Balcells for his comments on this paper prior to its submission.
We also wish to thank the anonymous referee for their comments and suggestions.

\clearpage

\figcaption[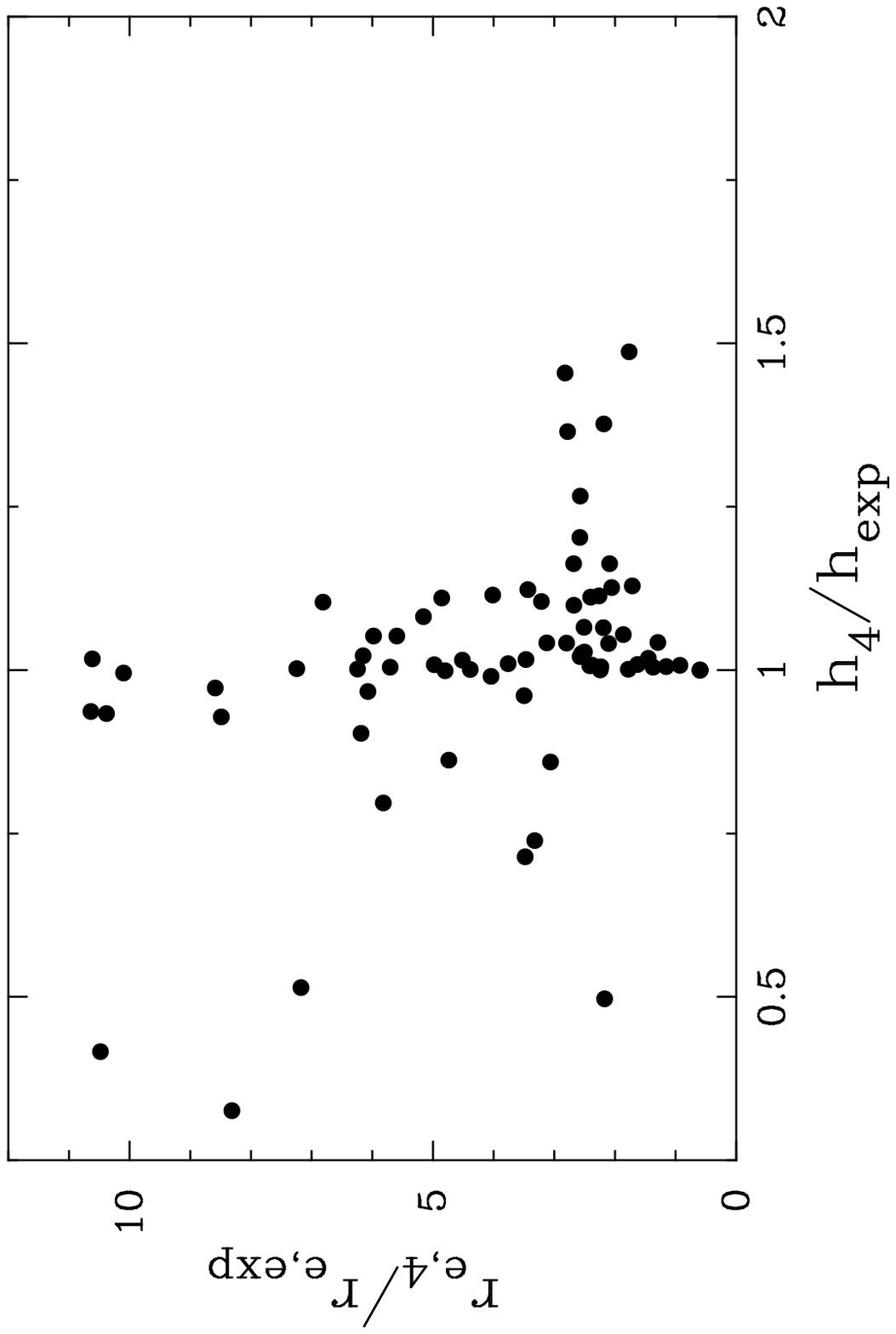]{For each galaxy, the value of $r_e$ from the $n$=4 
bulge model, divided by the value of $r_e$ from the $n$=1 bulge model, is 
plotted against the corresponding ratio of $h$ from the two associated 
exponential disk models.  While the scale-length of the exponential disk 
remains largely unchanged, the effective radius of the bulge is shown to 
be larger with an $r^{1/4}$ bulge model than with an exponential bulge model. 
(K-band data.)\label{4to1-fig}}

\clearpage

\begin{deluxetable}{rccc}
\tablewidth{0pt}
\tablecaption{Best-fitting bulge models\label{chi2-tab}}
\tablehead
{
\colhead{Morphological} & \multicolumn{3}{c}{No.\ of galaxies best fit with each model} \\[.2ex]
\colhead{Type} & \colhead{Exponential} & \colhead{($r^{1/2}, r^{1/4}$)} & \colhead{(no data)}
}
\startdata
\multicolumn{4}{c}{B-band} \\
  Sa,Sab,Sb &  10  &  12  &   0  \\
  Sbc,Sc,Scd &  31  &  16  &   1  \\
  Sd,Sdm,Sm & 11   &  2  &   0  \\
\tableline
\multicolumn{4}{c}{V-band} \\
  Sa,Sab,Sb &  7  &  8  &   7  \\
  Sbc,Sc,Scd & 24   &  18  &   6  \\
  Sd,Sdm,Sm &  10  &  2  &   1  \\
\tableline
\multicolumn{4}{c}{R-band} \\
  Sa,Sab,Sb & 6   &  16  &   0  \\
  Sbc,Sc,Scd &  25  & 21   &   2  \\
  Sd,Sdm,Sm &  10  & 3  &   0  \\
\tableline
\multicolumn{4}{c}{I-band} \\
  Sa,Sab,Sb &  5  &  11  &   6  \\
  Sbc,Sc,Scd & 21   &  24  &   3  \\
  Sd,Sdm,Sm & 10   &  3  &   0  \\
\tableline
\multicolumn{4}{c}{H-band} \\
  Sa,Sab,Sb &  6  &  8  &   8  \\
  Sbc,Sc,Scd & 13   &  5 &   30  \\
  Sd,Sdm,Sm &  4  &  2  &   7  \\
\tableline
\multicolumn{4}{c}{K-band} \\
  Sa,Sab,Sb &  8  &  13  &   1  \\
  Sbc,Sc,Scd & 27   &  18  &   3  \\
  Sd,Sdm,Sm &  12  &  1  &   0  \\
\enddata                         
\tablecomments{This table lists the number of spiral galaxy bulges
whose surface brightness profiles are best fit with an 
exponential model, and the number of galaxy bulges that are better 
modelled with the alternative ($r^{1/2}$ or $r^{1/4}$) profile. 
The fourth column displays the number of 
galaxies for which no $\chi ^2$ goodness-of-fit estimate was available.  
$\chi ^2$ data taken from de Jong 
(\url{http://cdsweb.u-strasbg.fr/htbin/Cat?J/A+AS/118/557}). 
%
%
}
\end{deluxetable}

\clearpage

\begin{deluxetable}{lcc}
\tablewidth{0pt}
\tablecaption{Comparison of the $r_{e}/h$ data distributions\label{stats-tab}}
\tablehead
{
\colhead{Band} & 
\colhead{$\left<\frac{r_e}{h}\right>_1-\left<\frac{r_e}{h}\right>_2$} 
& \colhead{Prob(t)} 
}
\startdata
\multicolumn{3}{c}{Exponential bulge model data} \\	  
\multicolumn{3}{c}{1:(S0,Sa,Sab,Sb) vs.\ 2:(Sbc,Sc,...Sm,Irr)} \\
K-band & (0.119-0.162)=--0.043  &  02\%  \\
R-band & (0.112-0.124)=--0.012   & 34\%  \\
 & & \\
\multicolumn{3}{c}{Exponential bulge model data} \\	  
\multicolumn{3}{c}{1:(Sa,Sab,Sb) vs.\ 2:(Sd,Sdm,Sm)} \\
K-band & (0.118-0.195)=--0.077 &  03\%   \\
R-band & (0.111-0.126)=--0.015 &  52\%  \\
 & & \\
\multicolumn{3}{c}{Best-fitting bulge models} \\        
\multicolumn{3}{c}{1:(Sa,Sab,Sb) vs.\ 2:(Sd,Sdm,Sm)} \\
K-band & (0.240-0.228)=+0.012  &  79\% \\
H-band & (0.485-0.173)=+0.312  &  07\% \\
I-band & (0.348-0.190)=+0.158  &  24\% \\
R-band & (0.333-0.186)=+0.147  &  23\% \\
V-band & (0.671-0.161)=+0.510  &  14\% \\
B-band & (0.321-0.160)=+0.161  &  18\% \\
\enddata
%
%
\tablecomments{Comparison of the $r_e/h$ data distributions for different 
morphological-type bins.  Column 1 shows the passband used.  The difference 
between the mean values from the two distributions (as listed 1: and 2: 
in the table sub-headings) is shown in column 2.
Column 3 gives the significance that the two distributions have 
the same mean value, as derived from Student's t-test and allowing
for different population variances between the two data sets. 
Small probabilities indicate that the data sets are different.}

\end{deluxetable}

\end{document}